\documentclass[journal]{IEEEtran}
\usepackage[utf8]{inputenc}

\ifCLASSINFOpdf
  \usepackage[pdftex]{graphicx}
  \graphicspath{{../} {../pdf/} {../jpeg/}}
  \DeclareGraphicsExtensions{.pdf,.jpeg,.png}
\else
  \usepackage[dvips]{graphicx}
  \graphicspath{{../} {../eps/}}
  \DeclareGraphicsExtensions{.eps}
\fi

\usepackage{hyperref} 
\usepackage{amssymb}
\usepackage{amsmath}
\usepackage{amsthm}
\usepackage{bm}
\usepackage{float}
\usepackage{color}
\usepackage{colortbl}
\usepackage{scalerel,stackengine}
\usepackage{multicol}
\usepackage{multirow}
\usepackage{graphicx}
\usepackage{diagbox}
\usepackage{multirow}
\usepackage{tabularx}
\usepackage[flushleft]{threeparttable}
\usepackage{stfloats}
\usepackage{subfigure}
\usepackage{setspace}

\usepackage{algorithm}
\usepackage{algpseudocode}

\makeatletter
\newenvironment{Method}[1][htb]{%
    \renewcommand{\ALG@name}{Method}
   \begin{algorithm}[#1]%
  }{\end{algorithm}}
\makeatother

\title{A Comparative Study of Polynomial Chaos Expansion-Based Methods for Global Sensitivity Analysis in Power System Uncertainty Control}

\author{
{Xiaoting Wang, Rong-Peng Liu, Xiaozhe Wang,  Fran\c{c}ois Bouffard}
 \thanks{ This work was supported partially by Natural Sciences and Engineering
 Research Council (NSERC) Discovery Grant, NSERC RGPIN-2022-03236 and partially by the Fonds de Recherche du Qu\'{e}bec-Nature et technologies under Grant FRQ-NT 320645.}
}
\date{January 2022}

\begin{document}
\maketitle

\begin{abstract}
In this letter, we compare three polynomial chaos expansion (PCE)-based methods for \color{black} ANCOVA (ANalysis of COVAriance) indices based \color{black} global sensitivity analysis for correlated random inputs in two power system applications. 
Surprisingly, the PCE-based models built  
with independent inputs after decorrelation 
may not give the most accurate \color{black} ANCOVA 
\color{black} indices, \color{black}
though this approach seems to be the most correct one and was applied in \cite{Caniou2012global} in the field of civil engineering.  
In contrast, the PCE model built using correlated random inputs directly yields the most accurate \color{black} ANCOVA  
indices for global sensitivity analysis\color{black}. Analysis and discussions about the errors of different PCE-based models will also be presented.  
These results provide important guidance for uncertainty management and control in power system operation and security assessment.
\end{abstract}


%
\vspace{-0.165in}
\section{Introduction}
\vspace{-0.04in}
As the penetration of intermittent renewable generation increases, 
global sensitivity analysis (GSA) has been explored to quantify the impacts of inputs uncertainties on 
the system model responses (e.g., voltage variation). 
\color{black}
While Monte Carlo (MC) simulations are commonly used to calculate Sobol' indices, the high computational time makes it impractical \cite{Okten2021}. Spectral techniques \cite{Liu2015,Mara2009}   
have been exploited for GSA, which, nevertheless, may work only for low-dimensional problems (e.g., the Fourier amplitude sensitivity test  \cite{Liu2015}) or independent random inputs (e.g., the random balance design \cite{Mara2009}). Meanwhile, surrogate model-based GSA has been adopted due to its high efficiency and accuracy \cite{Ye2021,Ni2018}. 
\color{black}
In \cite{Ye2021}, Ye et al. developed a kriging-based surrogate model to calculate Sobol' indices and to  quantify the impact of \textit{independent} stochastic power injections on voltage variations without considering correlations. 
Polynomial chaos expansion (PCE), 
another popular surrogate model, has been widely applied in power system uncertainty quantification (e.g., \cite{Xu2018,Xu2019,Zhang2013}). 
Particularly, to consider \textit{correlated} random inputs, 
Ni et al. applied a PCE-based model 
to calculate global sensitivity 
indices 
in power flow problems \cite{Ni2018}, 
i.e.,  evaluate how the correlated random 
power injections affect 
the bus voltage magnitudes.  
However, whether the PCE-based model was built with or without decorrelation was not discussed, even though either method could introduce errors in the calculated global sensitivity indices\color{black}.  

In fact, to the best of our knowledge, there have been no 
investigations or comparisons \color{black} 
regarding the use of PCE-based models in estimating global sensitivity indices for correlated random inputs in power systems, \color{black} which nevertheless are crucial for effective uncertainty control. \color{black} In previous works, Mara et al. \cite{Mara2021} 
calculated two  sensitivity indices based on  
ANOVA \color{black} (Analysis of variance) techniques for dependent inputs, which, nevertheless, may be time-consuming and impractical for high-dimensional problems.
\color{black} Another two PCE-based methods were suggested to handle correlated random inputs in GSA \color{black} based on 
ANCOVA (ANalysis of
COVAriance). The first method involves building a PCE model from independent random inputs, which ensures the orthogonality of polynomial bases and convergence of the PCE model in the $L_2$ norm (e.g., \cite{Caniou2012global}, \cite{Liu2021data}). As such, the first method will handle the correlated random inputs by decorrelations through the Nataf transform or the Rosenblatt transform 
(e.g., \color{black} \cite{Caniou2012global} 
in the context of  civil engineering\color{black}). \color{black}
However, it will be shown in this letter, decorrelation will inevitably introduce errors in estimating \color{black} ANCOVA
\color{black} indices. 
The second method, suggested in a previous study \cite{Torre2019Data}, involves ignoring dependency when building a PCE model, which can still yield reasonable accuracy, although theoretical proof was not provided. 

In this letter, we will compare the accuracy of different PCE-based models (without decorrelation, and with decorrelation using different nonlinear transforms) in estimating the \color{black} ANCOVA indices for global sensitivity analysis 
\color{black} in two power system applications based either on simulated or real-world data.  
Surprisingly, the PCE model built directly using correlated random inputs (ignoring the dependency) yields the most accurate \color{black}ANCOVA \color{black} indices in both applications. Analysis and discussions regarding the errors of different PCE-based models will also be presented. 
\color{black}By utilizing the obtained sensitivity information, effective mitigation measures are developed to reduce the variance of the system responses and enhance the system performance. 
\color{black}These results provide important guidance regarding uncertainty management and control design in practical power system applications. 

%
\vspace{-0.12in}
\section{The Polynomial Chaos Expansion-based Global Sensitivity Analysis Method}
\vspace{-0.05in}
\subsection{Covariance-Based Sensitivity Indices}
Consider a model  \small $Y=G(\bm{Z})$ \normalsize with a random input vector \small $\bm{Z} =\{Z_1,\cdots,Z_D\} \in\mathbb{R}^{D}$\normalsize,  where \small $\bm{Z}$ \normalsize could be volatile renewables (e.g., wind speed), load variations, etc.  
The system response \small$Y$\normalsize is also a random variable 
(e.g., voltage variation in \cite{Ye2021,Wang2022CPF,Bizzarri2020}, 
probabilistic total transfer capability (PTTC) in \cite{Wang2021,Liu2019}). 
Covariance-based sensitivity indices can be utilized to quantify how the variation in 
\small ${Z}_j$ \normalsize  
affects the variance 
of the system response \small$Y$\normalsize. Based on these indices, 
critical inputs can be  identified, 
and controls (smoothing out the critical inputs,  i.e., reducing their variance to zero) can be designed to reduce the variance of  \small$Y$\normalsize  and enhance the system performance in the probability sense. 
\color{black}As shown in \cite{Caniou2012global}, the model
\small $Y=G(\bm{Z})$ \normalsize can be decomposed as 
functions of \small$Z_j$\normalsize\color{black}:\color{black}
 \vspace{-0.08in}
\footnotesize
\begin{equation}
\setlength{\abovedisplayskip}{6pt}
\setlength{\belowdisplayskip}{2pt}
\label{eq:HDMR}
\begin{aligned}
\setlength{\jot}{-1\baselineskip}
&Y =G(\bm{Z}) =G_{0}  +  \sum_{1\leq{j} \leq D}G_{j}(Z_{j})+ \sum_{1\leq{j} < d \leq{D}}G_{j,d}(Z_{j},Z_{d}) + \cdots \\
&\sum_{1\leq{j_1}<\cdots<{j_{m}}\leq{D}} G_{j_1,\cdots,j_m}(Z_{j_1},\cdots,Z_{j_m}) 
 +\cdots  G_{1,\cdots,D}(Z_{1},\cdots,Z_{D})
\end{aligned}
\end{equation}
\normalsize
where \small $G_0$ \normalsize is the mean \small $\mathbb{E}[Y]$\normalsize; \small $G_j{(Z_j)}$ \normalsize  
represents the effect of  a single random input \small $Z_j$ \normalsize on \small $Y$\normalsize;  
\small $G_{j,d}(Z_{j},Z_{d})$ \normalsize
describes 
the interaction between 
random inputs $Z_{j}$ and $Z_{d}$ on $Y$, etc. 
Let \small $\mathcal{A} = \{1,\cdots,D\}$ \normalsize and $\beta$ be a subset of $\mathcal{A}$, then (\ref{eq:HDMR}) can be represented as
\footnotesize
\vspace{-0.02in}
\begin{equation}
\vspace{-0.092in}
\setlength{\abovedisplayskip}{-4pt}
\setlength{\belowdisplayskip}{-6pt}
\label{eq:HDMR_1}
\begin{aligned}
Y=G_{0} + \sum_{\beta\subseteq \mathcal{A}, \beta\not = \emptyset}G_{{\beta}}(\bm{Z}_{{\beta}}) 
\end{aligned}
\end{equation}
\normalsize
Now, let us write the variance of the model response $Y$: 
\footnotesize
\begin{equation}
\setlength{\abovedisplayskip}{0pt}
\setlength{\belowdisplayskip}{0pt}
\label{eq:ANCOVA}
\begin{aligned}
&\mathrm{Var}[Y] = \mathbb{E}\left[ (Y-\mathbb{E}[Y])^2\right] = \mathbb{E} \left[(Y-G_{0})\left(\sum_{\beta\subseteq \mathcal{A}}G_{\mathbf{\beta}}(\bm{Z}_{\mathbf{\beta}})\right)\right] \\
&= \mathrm{Cov}\left[Y,\sum_{\beta\subseteq \mathcal{A}}G_{\mathbf{\beta}}(\bm{Z}_{\mathbf{\beta}})\right] =\sum_{\beta\subseteq \mathcal{A}} \mathrm{Cov}\left[Y,G_{\mathbf{\beta}}(\bm{Z}_{\mathbf{\beta}})\right]
\end{aligned}
\end{equation}
\normalsize
Note that the second to the last equality 
is based on the definition of covariance \footnotesize $\mathrm{Cov}[X_1,X_2]:=\mathbb{E}[(X_1-\mathbb{E}[X_1])(X_2-\mathbb{E}[X_2])]$. \normalsize The last equality 
is based on the property of covariance. 
\normalsize Substituting the expression of $Y$ in (\ref{eq:HDMR_1}) to  
\normalsize 
(\ref{eq:ANCOVA}), we have: 
\footnotesize
\begin{equation}
\setlength{\abovedisplayskip}{0pt}
\setlength{\belowdisplayskip}{0pt}
\label{eq:ANCOVA_cov}
\begin{aligned}
&\mathrm{Var}[Y] 
= \sum_{\beta\subseteq \mathcal{A}}\left[ \mathrm{Var}\left[G_{\mathbf{\beta}}(\bm{Z}_{\mathbf{\beta}})\right]+ \mathrm{Cov}\left[G_{\mathbf{\beta}}(\bm{Z}_{\mathbf{\beta}}), \sum_{\substack{v\subseteq \mathcal{A}\\ v\not = \emptyset,v\not = \beta}}G_{v}(\bm{Z}_{v})\right]\right]
\end{aligned}
\end{equation}
\normalsize
The above technique  to decompose 
$\mathrm{Var}[Y]$ 
is referred  as the ANCOVA decomposition 
\cite{Caniou2012global}. The decomposition in (\ref{eq:ANCOVA}) indicates that 
$\mathrm {Var}{[Y]}$ can be expressed as the sum of the covariance of \small $G_{\mathbf{\beta}}(\bm{Z}_{\mathbf{\beta}})$ \normalsize and the response $Y$ for all $\beta \subseteq \mathcal {A}$. Therefore, 
the ANCOVA index 
for a single random input $Z_j$, $j\in \beta$ can be defined as  $S_{j} = \frac{\mathrm{Cov}\left[Y,G_j({Z}_{{j}})\right]}{\mathrm{Var{[Y]}}}$, which depicts the total effect of the variation of $Z_{j}$ on the variance of $Y$. Using  \eqref{eq:ANCOVA_cov},  $S_{j} $ can be further separated into the uncorrelated and correlated effects of $Z_j$ on $\mathrm{Var}[Y]$: 
\footnotesize
\begin{equation}
\setlength{\abovedisplayskip}{0pt}
\setlength{\belowdisplayskip}{0pt}
\label{eq:ANCOVA_Indices}
\begin{aligned}
\setlength{\jot}{-1\baselineskip}
& S_{j}  =\frac{\mathrm{Cov}\left[Y,G_j({Z}_{{j}})\right]}{\mathrm{Var{[Y]}}}=S_{j}^{(\mathrm{U})}+S_{j}^{(\mathrm{C})} \vspace{-0.03in}\\ 
& S_{j}^{(\mathrm{U})} = \frac{\mathrm{Var}[G_{{j}}({Z}_{{j}})]}{\mathrm{Var}[Y]},  
 S_{j}^{(\mathrm{C})}= \frac{\mathrm{Cov}\left[G_{{j}}({Z}_{j}), \sum_{\substack{v\subseteq \mathcal{A}\\ v \not = \{j\}}}G_{v}(\bm{Z}_{v})\right]}{\mathrm{Var}[Y]}
\end{aligned}
\end{equation}
\normalsize
where \small $S_{j}^{(\mathrm{U})}$ \normalsize is the uncorrelated contribution of  \small $Z_j$ \normalsize on \small  $\mathrm{Var}[Y]$\normalsize,   
\small $S_{j}^{(\mathrm{C})}$ \normalsize is the correlated contribution of \small $Z_j$ \normalsize to \small $\mathrm{Var}[Y]$\normalsize. 
For example, if \small $Z_j, j=\{1,2\}$ \normalsize are the wind speeds of two wind farm sites 
and we want to study how the variation of the two wind speeds will affect the variance of the model response $Y$, the total transfer capability (TTC) in this case,  then \small $S_{1}^{(U)}$ \normalsize describes how 
$Z_1$ itself affects the TTC $Y$; \small $S_{1}^{(C)}$ \normalsize describes how the 
correlation between the two wind speeds affect $Y$; $S_1$ describes the total effect of the variation of the wind speed $Z_1$ on $Y$. Therefore, $Z_j$ with the largest $S_j$ are regarded as the dominant random inputs because reducing the variance of these inputs (smoothing them out) can reduce $\mathrm{Var}[Y]$  most effectively. 
MC simulations can be  carried out to identify the terms \small $G_j(Z_j)$, $G_{j,d}(Z_j,Z_d), ...$ \normalsize in (\ref{eq:HDMR}) as discussed in 
\cite{Caniou2012global}, which, nevertheless, is computationally expensive. In contrast, 
a PCE-based model can be leveraged to calculate efficiently the 
sensitivity indices  
of correlated random inputs.
\vspace{-0.16in}
\subsection{The PCE-based ANCOVA Method} \label{sec:ANCOVA}
Given a random vector $\bm{Z}$ with finite second-order moments, 
it was shown in \cite{Caniou2012global} that the stochastic response 
$Y=G(\bm{Z})$ with finite second-order moments 
can be approximated by a PCE-based model $G^\mathrm{pc}(\bm{Z})$ which 
is a 
series of orthogonal polynomials of 
$\bm{Z}$: 
\footnotesize
\begin{equation}
\setlength{\abovedisplayskip}{-1pt}
\setlength{\belowdisplayskip}{-1pt}
\label{eq:PCE}
    \begin{aligned}
    Y\approx G^\mathrm{pc}(\bm{Z}) = 
    \sum_{\bm{k}\in \mathbb{N}^{{D}} }\color{black}a_{\bm{k}} P_{\bm{k}}(\bm{Z}) = \bm{A}^{T}\bm{\mathcal{P}}(\bm{Z})
    \end{aligned}
\end{equation}
\normalsize
where \small $\bm{k}=\{k_j, j=1,\cdots,D\}$ \normalsize are multi-indices for $P_{\bm{k}}(\bm{Z})$. $a_{\bm{k}}$ are $L$ unknown coefficients to be determined. \small $\bm{A}=\{a_{\bm{k}_0},\cdots, a_{\bm{k}_{L-1}}\}^{T}$ \normalsize and \small $\bm{\mathcal{P}}(\bm{Z})=\{P_{\bm{k}_0}(\bm{Z}),\cdots, P_{\bm{k}_{L-1}}(\bm{Z})\}^{T}$ \normalsize are the coefficients and polynomial bases in vector forms. 
\color{black}
Particularly, $P_{\bm{k}}(\bm{Z})$ have to satisfy the orthogonal condition that \small $\int_{\Omega} P_{\bm{k}}({\bm{Z}})P_{\bm{m}}({\bm{Z}}){\rho}({\bm{Z}})d{\bm{Z}} = 0$ \normalsize for $\bm{k} \neq \bm{m}$, where $\Omega$ is the support of $\bm{Z}$ and $\rho(\bm{Z})$ is the joint probability density function (PDF) of $\bm{Z}$\color{black}, so that \small $ \mathbb{E} \left[ \left( Y- G^{\mathrm{pc}}(\bm{Z})\right)^2 \right]\rightarrow 0$ when $L\rightarrow +\infty$ \normalsize
according to Cameron-Martin theorem \cite{Cameron1947}. 

If we write the PCE-based model (\ref{eq:PCE}) in terms of the high dimensional model representation (HDMR) (\ref{eq:HDMR}), we have:  
\vspace{-0.1in}
\footnotesize
\begin{equation}
\setlength{\belowdisplayskip}{3pt}
\label{eq:HDMR_PCE_decom}
\begin{aligned}
&\hat{Y} =G^{\mathrm{pc}}(\bm{Z}) = 
G_{0}^{\mathrm{pc}}+ \sum_{1\leq {j} \leq D} G_{j}^{\mathrm{pc}}(Z_{j})+ \sum_{1\leq{j} < d \leq{D}}G_{j,d}^{\mathrm{pc}}(Z_{j},Z_{d}) \\
& +  \cdots+G_{1,\cdots,D}^{\mathrm{pc}}(Z_{1},\cdots,Z_{D}) 
\end{aligned}
\end{equation}
\normalsize
Comparing (\ref{eq:PCE}) and (\ref{eq:HDMR_PCE_decom}), we have 
\footnotesize
\vspace{-2pt}
\begin{equation}
\setlength{\abovedisplayskip}{4pt}
\setlength{\belowdisplayskip}{0pt}
\label{eq:HDMR_PCE_fun}
\begin{aligned}
\setlength{\jot}{-1\baselineskip}
G_{0}^{\mathrm{pc}} &=a_0 \\ 
 G_{j}^{\mathrm{pc}}(Z_{j}) &= \sum^{p}_{ k_j=1 \color{black}} a_{k_j} P_{k_j}(Z_j)
 \\
G_{j,d}^{\mathrm{pc}}(Z_{j},Z_{d}) &= \sum_{k_j=1 }^{p}\sum_{k_d=1 }^{p} a_{\bm{k}_{j,d}} P_{\bm{k}_{j,d}}(Z_j,Z_d)\\ 
\cdots 
\end{aligned}
\end{equation}
\normalsize
where $a_{0}$ is the constant term, and $p$ is the order of the PCE-based model, \color{black} determined by the stopping criteria in (23)-(24) of \cite{Wang2021}. \color{black} $G_{j}^{\mathrm{pc}}(Z_{j})$ includes the terms \small $P_{k_j}(Z_j)$ \normalsize  that depend only on $Z_j$, 
\small $G_{j,d}^{\mathrm{pc}}(Z_{j},Z_{d})$ \normalsize includes the terms \small 
 $P_{\bm{k}_{j,d}}(Z_j,Z_d)$ \normalsize depending only on $Z_j$ and $Z_d$, etc. 

Based on 
\eqref{eq:HDMR_PCE_decom} and \eqref{eq:HDMR_PCE_fun}, 
it can be seen that ANCOVA indices \eqref{eq:ANCOVA_Indices} can be calculated from the PCE-based model 
(\ref{eq:PCE}), which turns out to be much more efficient compared to MC simulations. 
Assuming there are $M_L$ samples of \small $\bm{Z}^{(l)}$, $l=1,...,M_L$ \normalsize
and 
corresponding ${\hat{Y}}^{(l)}$ are evaluated efficiently from the PCE-based model, i.e., \small  ${\hat{Y}}^{(l)}=G^\mathrm{pc}(\bm{Z}^{(l)})$\normalsize, then the sample mean \footnotesize
$\widehat{\mathbb{E}}[\hat{Y}]$ \normalsize and the sample variance \footnotesize$ \widehat{\mathrm{Var}}[\hat{Y}]$ \normalsize can be calculated.
The sample covariance of \small$Y$ \normalsize and \small$G_j(Z_j)$ \normalsize as well as the sample variance of \footnotesize $G^\mathrm{pc}_{j}(\bm{Z}_j^{(l)})$ \normalsize can also be estimated: 
\scriptsize
\scriptsize
\begin{subequations}
\setlength{\abovedisplayskip}{-1pt}
\setlength{\belowdisplayskip}{1pt}
\begin{align}
&\widehat{\mathrm{Cov}}\left[\hat{Y},G_j^{\mathrm{pc}}({Z}_{{j}})\right]=\frac{1}{M_L-1}\sum_{l=1}^{M_\mathrm{L}}\left[\hat {{Y}}^{(l)}-\widehat{\mathbb{E}}[\hat{Y}]
\right]\left[G^\mathrm{pc}_{j}(\bm{Z}_j^{(l)})-\widehat{\mathbb{E}}[G^\mathrm{pc}_{j}(Z_j)] 
\right]\nonumber \\ 
&\mbox{\normalsize where \scriptsize} \widehat{\mathbb{E}}[G^\mathrm{pc}_{j}(Z_j)]=\frac{1}{M_\mathrm{L}}\sum_{l =1}^{M_\mathrm{L}} G^\mathrm{pc}_j(\bm{Z}_j^{(l)})\label{eq:sample cov} \\ 
&\widehat{\mathrm{Var}}[G^\mathrm{pc}_{j}(Z_j)]=\frac{1}{M_L-1}\sum_{l=1}^{M_\mathrm{L}} \left[G^\mathrm{pc}_{j}(\bm{Z}_j^{(l)})-\widehat{\mathbb{E}}[G^\mathrm{pc}_{j}(Z_j)]\right]^2\label{eq:sample var}
\end{align}
 \end{subequations}
\normalsize
As a result, 
ANCOVA indices in (\ref{eq:ANCOVA_Indices}) can be
obtained:\color{black}
\vspace{-0.06in}
\begin{equation}
\setlength{\abovedisplayskip}{5pt}
\setlength{\belowdisplayskip}{1pt}
\label{eq:ANCOVA_PCE}
\scriptsize
S_{j}=\frac{\widehat{\mathrm{Cov}}\left[\hat{Y},G_j^{\mathrm{pc}}({Z}_{{j}})\right]}{\widehat{\mathrm{Var}}[\hat{Y}]\color{black}},  
 S_{j}^{(\mathrm{U})} = 
 \frac{\widehat{\mathrm{Var}}[G^\mathrm{pc}_{j}({Z_j)]}}
{\widehat{\mathrm{Var}}[\hat{Y}\color{black}]}, 
 S_{j}^{(\mathrm{C})}= S_{j} - S_{j}^{(\mathrm{U})}
\end{equation}
\normalsize
\noindent 

Certainly, to calculate $S_j$ in \eqref{eq:ANCOVA_PCE}, the PCE model \eqref{eq:PCE}  should be built first. If random inputs $\bm{Z}$ are mutually \textit{independent}, \small $P_{\bm{k}}(\bm{Z}) $ \normalsize  can be constructed
by the tensor product of the univariate orthogonal polynomial bases \small $\phi^j_{k_{j}}(Z_{j})$ \normalsize to ensure the orthogonality of \small 
 $P_{k}(\bm{Z})$, \normalsize i.e.,  \small $P_{\bm{k}}(\bm{Z}) 
 = \prod_{j=1}^{D}\phi^j_{k_j}(Z_{j})$\normalsize, where \color{black}
\small $\phi^j_{k_{j}}(Z_{j})$ \normalsize can be first calculated from raw data or an assumed probabilistic model of $Z_j$, and $k_{j}$ is the corresponding degree of \small  $\phi^j_{k_{j}}(Z_{j})$\normalsize. \color{black} \color{black} Specially, in this letter the moment-based method (see Section III-B in \cite{Wang2021}) is applied to construct univariate polynomial bases \small $\phi^j_{k_{j}}(Z_{j})$\normalsize.  
After \small $P_{\bm{k}}(\bm{Z})$ \normalsize is constructed, \color{black}
$a_{\bm{k}}$ can be calculated by advanced regression methods, e.g., least angle regression (LAR), using 
$M_p$ (a small number) sample evaluations. 
Please refer to 
\cite{ Wang2021} 
for more details. 
Since $\bm{Z}$ has mutually \textit{independent} random factors, \small  $\mathbb{E}[Y]$\normalsize, 
 \small $\mathrm{Var}[Y]$\normalsize, and $S_j$ can be directly obtained from the coefficients of the PCE-based model \eqref{eq:PCE} \color{black}(i.e., Sobol' indices \cite{Ye2021,Sudret2008}) \color{black} or through sample evaluations using \eqref{eq:PCE} and \eqref{eq:ANCOVA_PCE}.

However, if inputs $\bm{Z}$ (e.g., wind speeds) are \textit{correlated}, which is true in general, the multivariate polynomial bases $P_{k}(\bm{Z})$, 
\textit{cannot} be  constructed \color{black}  purely \color{black} through the tensor product of the  univariate polynomial bases $\phi^j_{k_{j}}(Z_{j})$ 
because the sufficient condition for the convergence of the PCE model that 
$P_{\bm{k}}(\bm{Z})$ are orthogonal with respect to the joint PDF of $\bm{Z}$ is not satisfied \cite{Soize2004}.   
\color{black} 
In previous literature, two methods were suggested to handle the correlated random inputs \color{black} for ANCOVA indices, \color{black} yet no theoretical proof was given:
1) building a PCE model by ignoring the input dependency as \cite{Torre2019Data} claimed that a reasonably accurate response can still be achieved; 
2) building a PCE model after decorrelating correlated random inputs as \cite{Caniou2012global} claimed that a PCE model built with independent inputs still holds for correlated inputs with the same marginal distributions. 
To the best of our knowledge, no comparisons have been made inside or outside the power community, which in turn is critical for effective uncertainty management. In this letter, we will investigate and compare these two PCE-based methods in estimating ANCOVA indices and uncertainty control \color{black} for power systems\color{black}. Analysis of the errors introduced by the two methods will also be presented. 

 The \textit{first} method denoted as \textit{PCE\_correlate} 
is to construct a PCE model \eqref{eq:PCE} 
based on $[\bm{Z}_{{p}},\mathbf{Y}_{p}]$ 
by ignoring the input dependencies. 

The \textit{second} method  includes decorrelating 
correlated samples $\bm{Z}_{{p}}$ into independent samples $\bm{U}_{{p}}$ 
through the Nataf (or Rosenblatt) transform and constructing the PCE model \eqref{eq:PCE} using the sample pair $[ \bm{U}_{{p}}, \bm{Y}_p]$. 
For simplicity, this method using the Nataf and the Rosenblatt transforms are denoted by \textit{PCE\_NT} and \textit{PCE\_RT}, respectively. The Nataf transform is 
used  
when $\bm{Z}$ has a  Gaussian copula, 
while the Rosenblatt transform  
is adopted when $\bm{Z}$ has more a complex correlation (e.g., nonlinear or tail dependence) 
\cite{Caniou2012global}.

Once the PCE-based model \eqref{eq:PCE} is built by one of the aforementioned two methods, \small $M_L$ \normalsize samples (\small$M_L\gg M_p$\normalsize)  of \small $\bm{Z}^{(l)}, l=1,...,M_L$ \normalsize
can be substituted into the PCE-based model  to obtain 
corresponding responses \small$\hat{Y}^{(l)}$ 
 \normalsize efficiently. Then the ANCOVA indices \small$S_{j}$ \normalsize for each \small $Z_j$ \normalsize can be estimated by (\ref{eq:ANCOVA_PCE}), based on which effective control measures can be designed to reduce the variance of the system response \small $Y$ \normalsize in the most effective way. 
The detailed steps of the two PCE-based methods for ANCOVA indices estimation and uncertainty control are summarized in  \textbf{Method} \ref{Met:first_method} and \textbf{Method} \ref{Met:sec_method}.

\noindent\textit{Remark} \color{black}1. The details of \textbf{Step 2} including 
\small $\phi_{k_j}^{j}(Z_j)$ \normalsize  construction, \color{black}PCE order $p$ selection \color{black} and  
coefficients $a_{\bm{k}}$ calculation can be found in  Section  III-B-Section III-D of \cite{Wang2021}. \color{black}Particularly, the LAR algorithm is implemented to determine the optimal order of the PCE, using the corrected leave-one-out cross-validation error ($e_{\mathrm{cloo}}$) index as the stopping criterion (see equations (23)-(24) in \cite{Wang2021}). The algorithm starts with an initial value of $p$ and evaluates $e_\mathrm{{cloo}}$ for different orders, comparing the results until either a prescribed accuracy is achieved or no further improvement can be made. Based on the comparison, the algorithm selects the final order that yields the best results. 

\noindent\textit{Remark}  \color{black}2. 
The PCE-based ANCOVA indices calculation is much computationally cheaper 
compared to MC simulations. The small number of sample evaluations in \textbf{Step 1} takes most of the time consumption, while building the PCE-based model in \textbf{Step 2} and evaluating ANCOVA indices in \textbf{Step 3} takes negligible time. In contrast,  MC simulations require running \small $(D+1)\times M_{L}$ \normalsize simulations for \small $Y$ \normalsize and \small$G_{j}(Z_j)$ \normalsize in \textbf{Step 4}, which is much more computationally expensive since \small $M_L \gg M_p$. \normalsize A comparison between the computational time will be given in Section \ref{section: study_A}.

\color{black}
\setlength{\intextsep}{4pt}
\setlength{\textfloatsep}{0pt}
\begin{Method}
\footnotesize
\setlength{\leftskip}{0pt}
\begin{algorithmic}
\footnotesize
\color{black}
\State \textbf{Step 1.} 
Generate a set of input samples $\bm{Z}_p \in \mathbb{R}^{M_p \times D}$ of $D$ random inputs $\bm{Z}$ (e.g., wind speeds),  
use deterministic power system analysis tools to get the model response $\bm{Y}_p \in \mathbb{R}^{M_p}$  that correspond to the $\bm{Z}_p$ samples. 
Pass the set of sample-response pairs $[\bm{Z}_p,\bm{Y}_p]$ to \textbf{Step 2}.\\
\State \textbf{Step 2.} Construct the \textit{PCE\_{correlate}} model in 
\eqref{eq:PCE}:
\begin{itemize}
\setlength{\itemsep}{0pt}
\setlength{\parsep}{0pt}
\setlength{\parskip}{0pt}
\item[a)] Build the univariate polynomials $\phi^j_{k_j}(Z_{j})$ using the moment-based method (see (17) in \cite{Wang2021}); 
\item[b)] Construct the multivariate polynomials $\bm{P_k}(\bm{Z})$ through the tensor product of $\phi^j_{k_j}(Z_{j})$;
\item[c)] Calculate the 
coefficients $\bm{a_k}$ using 
LAR based on the $M_p$ sample pairs obtained in \textbf{Step 1}.
\end{itemize}
\State  \textbf{Step 3.} Obtain the terms
$G_{j}^{\mathrm{pc}}(Z_{j})$, $G_{j,d}^{\mathrm{pc}}(Z_{j},Z_{d})$, ..., in the HDMR \eqref{eq:HDMR_PCE_fun} from the  
\textit{PCE\_{correlate}} model constructed in \textbf{Step 2}. \\
\State \textbf{Step 4.} Acquire a large number of $M_L$ input samples $\bm{Z}^{(l)}$ to evaluate  ${\hat{Y}}^{(l)}=G^\mathrm{pc}(\bm{Z}^{(l)})$ by the \textit{PCE\_{correlate}} model constructed in \textbf{Step 2}, and calculate 
$\widehat{\mathrm{Cov}}\left[\hat{Y},G_j^{\mathrm{pc}}({Z}_{{j}})\right]$ and $\widehat{\mathrm{Var}}[G^\mathrm{pc}_{j}(\bm{Z}_j^{(l)})$] by (\ref{eq:sample cov})-(\ref{eq:sample var}). \\ 
\State \textbf{Step 5.} \footnotesize Calculate the ANCOVA indices $S_j$ by \eqref{eq:ANCOVA_PCE}. 
Identify the critical random inputs with the highest $S_j$ values.\\
\State \textbf{Step 6.} Uncertainty control:
smooth out the critical random inputs identified in \textbf{Step 5}, i.e., reducing the variance of the critical random inputs to zero by, e.g., energy storage systems.    
\end{algorithmic}
\small
\caption{\small{\textit{PCE\_{correlate}} ANCOVA indices estimation and uncertainty control}}
\label{Met:first_method}
\end{Method}
\normalsize
\setlength{\intextsep}{0pt}
\setlength{\textfloatsep}{0pt}
\begin{Method}
\footnotesize
\begin{algorithmic}
\State \textbf{Step 1.} \color{black} The same as \textbf{Step 1} in \textbf{Method 1}. \color{black}
\\
\State \textbf{Step 2.} Construct the \textit{PCE\_{NT}} or \textit{PCE\_{RT}}  model in 
\eqref{eq:PCE}:
\begin{itemize}
\setlength{\itemsep}{0pt}
\setlength{\parsep}{0pt}
\setlength{\parskip}{0pt}
\item [a)] Decorrelate the input samples $\bm{Z}_p$ to $\bm{U}_p$ using the Nataf or Rosenblatt transform. Pass the data set $[\bm{U}_p,\bm{Y}_p]$ to \textbf{Step 2 b)}; 
\item [b)]Build the univariate polynomials $\phi^j_{k_j}(U_{j})$ using the moment-based method (see (17) in \cite{Wang2021}); 
\item [c)] Construct the multivariate polynomials $\bm{P_k}(\bm{U})$ through the tensor product of $\phi^j_{k_j}(U_{j})$;
\item[d)] Calculate the 
coefficients $\bm{a_k}$ using  
LAR based on the $M_p$ sample pairs obtained in \textbf{Step 2 a)}. 
\end{itemize} \\
\State\textbf{Step 3.} 
\footnotesize Obtain the terms $G_{j}^{\mathrm{pc}}(Z_{j})$, $G_{j,d}^{\mathrm{pc}}(Z_{j},Z_{d})$, ..., in the HDMR \eqref{eq:HDMR_PCE_fun} from the  
\textit{PCE\_{NT}} or \textit{PCE\_{RT}}  model constructed in \textbf{Step 2}. \\
\State \textbf{Step 4, 5, 6} are the same as those in \textbf{Method 1}. 
\end{algorithmic}
\small
\caption{\small{\textit{PCE\_{NT}} or \textit{PCE\_{RT}} ANCOVA indices 
estimation and uncertainty control}}
\label{Met:sec_method}
\end{Method}
\normalsize

 \color{black} At first glance, the \textit{second} method seems to be better as it considers the correlation between random inputs. The sufficient condition for the convergence of the PCE model that $P_{\bm{k}}$ are orthogonal with respect to the joint PDF of $\bm{Z}$ is satisfied in the \textit{second} method.  
However, the results in Section \ref{sec: Simulation} will show that the \textit{first} method typically gives a more accurate PCE-based model and thus more accurate estimations for ANCOVA indices. \color{black} The potential reasons and the analysis of errors are also discussed in Section
\ref{sec:discussion}. 
\normalsize

%

%
\vspace{-0.18in}
\section{Simulation Studies}\label{sec: Simulation}
\vspace{-0.05in}
We compare the estimations of ANCOVA indices using \textit{PCE\_correlate} (the \textit{first} method), \textit{PCE\_NT} and \textit{PCE\_RT} (the \textit{second} method), respectively, in two power system applications.  
In the first one,
we estimate the ANCOVA indices \color{black}in the probabilistic total transfer capability (PTTC) assessment, \color{black} 
aiming to find the 
\color{black} critical random inputs $Z_j$ dominating \color{black}  
the variance of PTTC. 
In this case, $\bm{Z}$ are generated from known distributions with linear correlations; $Y$ (PTTC) is Gaussian-like. \color{black} Refer to \cite{Wang2021} Section II for the formulation of PTTC assessment.
The second case \color{black}  presents the ANCOVA indices estimation for the 
economic dispatch (ED) problem considered in \cite{Wang2021ED}. 
In contrast to the first case, we use real-world data from the NREL’s Western Wind Data Set \cite{NREL2020}, which 
exhibits unknown distribution types and potentially complicated correlations.  To make matters more challenging,
$Y$ (the ED cost) is multimodal. \color{black} See Appendix in \cite{Wang2021ED} for the formulation of ED problem. The UQLab toolbox is adopted to build the PCE-based models \cite{UQdoc_11_104} \cite{UQdoc_13_106}\color{black}.
\color{black} 
\vspace{-0.21in}
\subsection{Case 1: Available Transfer Capability (ATC) Enhancement}\label{section: study_A}
\vspace{-0.056in}
Simulations are performed on the modified IEEE 24-bus reliability test system.
There are 6 random inputs (3 wind and 3 solar farms). 
\color{black} Response $Y$ considered is 
the  PTTC
\color{black} defined \color{black} from generators at bus 7 to loads at bus \{3,4,9\}. \color{black} The readers can find the system  configuration details in \color{black}
\small \url{https://github.com/TxiaoWang/PCE-based-ANCOVA.git}. \normalsize 
Sixty  
(\small $M_p = 60$\normalsize) 
sample pairs (\small $[\bm{Z}_{{p}}, \bm{Y}_p]$ \normalsize in the \textit{first} method or \small $[ \bm{U}_{{p}}, \bm{Y}_p]$ \normalsize in the \textit{second} method) are used to build the three PCE-based models (\textbf{Step 1-2} described in \textbf{Method} \ref{Met:first_method} and \textbf{Method} \ref{Met:sec_method}). 
\color{black}
The order $p$ 
determined by the LAR algorithm ((23)-(24) in \cite{Wang2021}) is 2 for all three PCE-based models. \color{black} 
The implementation of the LAR algorithm for selecting $p$  ensures fair comparisons among the models. \color{black} 
\color{black} 
Then \small $M_L = 10,000$ \normalsize samples of correlated random inputs are evaluated using the three PCE-based models (\textbf{Step 4})\color{black}. 
Fig. \ref{fig:CDFPDF_Comp_case24}-(a) 
shows that \textit{PCE\_correlate} (black) gives the closest results to the benchmark \color{black} latin hypercube sampling (LHS)-based \color{black} MC simulations (black) in estimating the cumulative distribution function (CDF). 
\begin{table}[!ht]
\vspace{-0.085in}
\setlength{\abovecaptionskip}{-0.17cm}
\setlength{\belowcaptionskip}{-0.4cm}
\caption{ 
ANCOVA indices  $S_j$ for PTTC from the three PCE-based models} 
\label{tab:ANCOVA_comp}
\renewcommand{\arraystretch}{1.0}
\centering
\resizebox{.96\columnwidth}{!}{
\begin{tabular}{c|c|c|c|c|c|c}
\hline
Input                     & $Z_1$  & $Z_2$  & $Z_3$  & $Z_4$  & $Z_5$  & $Z_6$  \\ \hline
{$\textit{PCE\_correlate}$} & 0.0257 & 0.0000 & 0.1887 & $\bm{0.2610} $& $\bm{0.2512}$ & $\bm{0.2735}$\\ \hline
$\textit{PCE\_NT}$        & 0.0437 & 0.0000 & 0.1548 & $\bm{0.4499}$ & 0.1896 & 0.1620 \\ \hline
$\textit{PCE\_RT}$        & 0.0179 & 0.0000 & 0.1205 & 0.2586 & 0.1523 & $\bm{0.3419} $\\ \hline
\end{tabular}}
\end{table}
\arrayrulecolor{black}
\begin{table}[htbp]
\vspace{-0.062in}
\footnotesize
\setlength{\abovecaptionskip}{-0.14cm}
\setlength{\belowcaptionskip}{-0.42cm}
\renewcommand{\arraystretch}{1.05}
\caption{\color{black}Statistics of PTTC before and after smoothing $Z_j$ by the benchmark MC simulations and the three PCE-based models}
\centering
\label{tab:smoothed_stat}
\resizebox{\linewidth}{!}{
\begin{tabular}{c|c|c|c|c|c|c|c} 
\hline
\multirow{2}{*}{\color{black}Methods}  & \multicolumn{2}{c|}{\color{black}$Z_4$}                                                                                                                                                                                                                                                        & \multicolumn{2}{c|}{\color{black}$Z_5$}                                                                                                                                                                                                                                                        & \multicolumn{2}{c|}{\color{black}$Z_6$}                                                                                                                                                                                                                                      & \multirow{2}{*}{\begin{tabular}[c]{@{}c@{}}\color{black}Before \\ 
\color{black}$\widehat{\sigma}[{Y}]$ \end{tabular}}   \\ 
\cline{2-7}
                          & \color{black} $\widehat{\sigma}[{Y}]$ & 
                          $\color{black}\Delta {\sigma}_{\mathrm{rr}}\%$& \color{black} $\widehat{\sigma}[{Y}]$  & 
                          $\color{black}\Delta {\sigma}_{\mathrm{rr}}\%$
                          &\color{black} $\widehat{\sigma}[{Y}]$ & 
                      $\color{black}\Delta {\sigma}_{\mathrm{rr}}\%$&                                                                                                                                                    \\ 
\hline
$\color{black}\textit{PCE\_correlate}$ & \color{black}0.7706                                                                & \color{black}$\bm{-2.11} $                                                                                                                                                                                                   & \color{black} 0.7785                                                                & \color{black}$\bm{-1.73} $                                                                                                                                                                                                 & \color{black}$0.7572$                                                                & \color{black}$\bm{-2.17} $                                                                                                                                                                                                   &\color{black} 0.9494                                                                                                                                             \\ 
\hline
$\color{black}\textit{PCE\_NT}$        & \color{black}0.7692                                                                & \color{black}$-2.29$                                                                                                                                                                                                    & \color{black}0.9605                                                                & \color{black}$21.24 $                                                                                                                                                                                                   & \color{black}0.9764                                                                & \color{black}$26.15$                                                                                                                                                                                                    & \color{black}1.1192                                                                                                                                             \\ 
\hline
$\color{black}\textit{PCE\_RT}$        & \color{black}0.9147                                                                & \color{black}$16.20$                                                                                                                                                                                                    & \color{black}1.0052                                                                & \color{black}$26.89$                                                                                                                                                                                                      & \color{black}0.7491                                                                & \color{black}$-3.22$                                                                                                                                                                                                    &\color{black} 1.1336                                                                                                                                             \\
\hline
\color{black}MC                        &\color{black} 0.7872                                                                & \color{black}--                                                                                                                                                                                                    & \color{black}0.7922                                                                & \color{black}--                                                                                                                                                                                                  & \color{black}$0.7740$                                                                & \color{black}--                                                                                                                                                                                                    &\color{black} $0.9702$                                                                                                                                             \\ 
\hline
\end{tabular}}
\vspace{-0.02in}
\begin{tablenotes}
\item\footnotesize{* \color{black} {
$\Delta {\sigma}_{\mathrm{rr}}= {(\hat{\sigma}[Y_{\mathrm{pc}}] - \hat{\sigma}[Y_{\mathrm{mc}}])}/{\hat{\sigma}[Y_{\mathrm{mc}}]}$} 
describes the normalized standard deviation estimation error 
by the three PCE models. } 
\end{tablenotes}
\vspace{-0.03in}
\end{table}
\normalsize	
\begin{table*}[hb]
\vspace{-0.29in}
\scriptsize
\setlength{\abovecaptionskip}{-0.16cm}
\setlength{\belowcaptionskip}{-0.5cm}
\renewcommand{\arraystretch}{1.0}
\caption{\color{black} \footnotesize Case 2: Comparison of the standard deviation of the ED cost before and after smoothing the three sets of top 40 dominant inputs by the benchmark MC simulations, and the three PCE-based models. } 
\label{tab:statistic_comp_case118}
\centering
\resizebox{\linewidth}{!}{
\begin{tabular}{c|ccc|ccc|ccc|c}
\hline
\multirow{2}{*}{Methods}                 & \multicolumn{3}{c|}{Set 1: Top 40s}                                                                                                        & \multicolumn{3}{c|}{Set 2: Top 40s}                                                                                                   & \multicolumn{3}{c|}{Set 3: Top 40s}                                                                                                  & \multirow{2}{*}{\begin{tabular}[c]{@{}c@{}}Before\\ $\widehat{\sigma}[{Y}]$ \end{tabular}} \\ \cline{2-10}
                                         & \multicolumn{1}{c|}{$\widehat{\sigma}[Y]$} & \multicolumn{1}{c|}{\color{black} $\Delta \sigma[Y] \%$} & \color{black} $\Delta \sigma_{\mathrm{re}}\%$ & \multicolumn{1}{c|}{$\widehat{\sigma}[Y]$} & \multicolumn{1}{c|}{\color{black}$\Delta \sigma[Y] \%$} & \color{black}$\Delta \sigma_{\mathrm{re}}\%$ & \multicolumn{1}{c|}{$\widehat{\sigma}[Y]$} & \multicolumn{1}{c|}{\color{black}$\Delta \sigma[Y] \%$} &\color{black} $\Delta 
                                        \sigma_{\mathrm{re}}\%$ &                                                                                               \\ \hline
\textit{PCE\_correlate} & \multicolumn{1}{c|}{$1.4925\times 10^{3}$   }                    & \multicolumn{1}{c|}{\color{black}$ -\bm{96.83}$}                    & \color{black}$ -\bm{0.0093}$                   & \multicolumn{1}{c|}{$6.0660 \times 10^{3}$ }                    & \multicolumn{1}{c|}{\color{black}$-\bm{87.12}$ }               & \color{black}$-\bm{0.0020}$                  & \multicolumn{1}{c|}{$1.0652 \times 10^{3}$ }                    & \multicolumn{1}{c|}{\color{black}$-\bm{97.74} $}              & \color{black}$-\bm{0.0190} $                 & $4.7090 \times 10^{4}$                                                                                    \\ \hline
\textit{PCE\_NT}        & \multicolumn{1}{c|}{$1.0467\times 10^{4}$}                    & \multicolumn{1}{c|}{\color{black}$-82.98$}                    & \color{black}$-0.1422$                    & \multicolumn{1}{c|}{$1.1919\times 10^{3}$}                    & \multicolumn{1}{c|}{\color{black}$-97.98$}               & \color{black}$0.1270$                   & \multicolumn{1}{c|}{$1.0586\times 10^{4}$}                    & \multicolumn{1}{c|}{\color{black}$-82.09$}              & 
\color{black}$-0.1442$& $5.9115 \times 10^4$                                                                                        \\ \hline
\textit{PCE\_RT}        & \multicolumn{1}{c|}{$6.9201 \times 10^{2}$}                    & \multicolumn{1}{c|}{\color{black}$-98.56$}                    &\color{black} $0.1194$                   & \multicolumn{1}{c|}{$5.1244 \times 10^{2}$}                    & \multicolumn{1}{c|}{\color{black}$-98.93 $}                & \color{black}$0.1379 $                   & \multicolumn{1}{c|}{$9.2717 \times 10^{2}$}                    & \multicolumn{1}{c|}{\color{black}$-98.07 $}              & \color{black}$0.0224$                   & $4.7920 \times 10^{4}$                                                                                       \\ \hline
MC                                       & \multicolumn{1}{c|}{$1.9115 \times 10^{3}$}                    & \multicolumn{1}{c|}{\color{black}$\bm{-95.94}$}                           & \color{black}--                   & \multicolumn{1}{c|}{$6.1449 \times 10^{3}$}                    & \multicolumn{1}{c|}{\color{black}$\bm{-86.94}$ }                     & \color{black}{--}                   & \multicolumn{1}{c|}{ $1.9205\times 10^{3}$}                    & \multicolumn{1}{c|}{\color{black}$\bm{-95.92 }$}                     & \color{black}--                 & $ 4.7069 \times 10^{4}$                                                                                      \\ \hline
\end{tabular}}
\vspace{-0.005in}
\begin{tablenotes}
\vspace{-1pt}
\item * Set 1, Set 2, and Set 3 are the three sets of top 40 inputs from \textit{PCE\_correlate}, \textit{PCE\_NT} and  \textit{PCE\_RT}, respectively. \color{black} $\Delta \sigma[Y] = (\hat\sigma{[Y_{\mathrm{after}}]}-\hat\sigma{[Y_{\mathrm{before}}]})/\hat\sigma{[Y_{\mathrm{before}}]}
$; $\hat\sigma{[Y_{\mathrm{before}}]}$  and $\hat\sigma{[Y_{\mathrm{after}}]}$ denote the estimated standard deviation of $Y$ before and after smoothing, respectively; $\Delta \sigma_{\mathrm{re}}= {(\Delta \sigma[Y_{\mathrm{pc}}]-\Delta \sigma[Y_{\mathrm{mc}}])}/{\Delta \sigma[Y_{\mathrm{mc}]}}$, describing how close the standard deviation reduction by the PCE-based model is to the one by the benchmark MC simulation. \color{black}
\end{tablenotes}
\normalsize
\end{table*}
\normalsize

Subsequently, 
ANCOVA indices \small $S_j$ \normalsize are calculated (\textbf{Step 5}) \color{black} 
and presented in 
Table 
\ref{tab:ANCOVA_comp}, which indicates that dominant inputs from the three PCE-based models are varying. 
\color{black}For validation, we smooth out \color{black} \small $Z_4$\normalsize, \small $Z_5$\normalsize,  and \small $Z_6$ \normalsize (\textbf{Step 6}) 
one by one to see how \small $\mathrm{Var}[\mathrm{PTTC}]$ \normalsize changes through \color{black}the PCE-based models and \color{black} MC simulations. \color{black} 
Table \ref{tab:smoothed_stat} shows that the \textit{PCE\_correlate} provides the \color{black} closest results to the benchmark MC simulations\color{black}. 
\color{black}
Clearly, smoothing \color{black}  \small $Z_4$\normalsize, \small $Z_5$\normalsize, and \small $Z_6$ \normalsize have comparable influences to \small $\mathrm{Var}[\mathrm{PTTC}]$\normalsize, yet smoothing \small$Z_6$ \normalsize is slightly more effective to \color{black} reduce \small $\mathrm{Var}[\mathrm{PTTC}]$ \normalsize (\color{black}$\approx 20\%$ reduction of $\hat{\sigma}[Y]$, i.e., the standard deviation of PTTC),
\color{black} thus enhancing the ATC. \color{black}
\color{black} These results \color{black} have verified the accuracy of 
\color{black}  the \color{black} ANCOVA \color{black} 
indices estimated by  
\textit{PCE\_correlate} shown in Table \ref{tab:ANCOVA_comp}.  
\color{black}
\color{black} Regarding the efficiency, \textit{PCE\_correlate} requires  $421.4$s for the ANCOVA indices estimation and is slightly faster ($\approx$ 2s) than the other two PCE-based models.  In contrast, the MC simulations take 490709.3s for ANCOVA indices calculations. 
\vspace{-0.21in}
\subsection{Case 2: Stochastic Economic Dispatch Problem}
\begin{figure}[!ht]
\vspace{-0.055in}
\setlength{\abovecaptionskip}{-0.21cm}
\setlength{\belowcaptionskip}{-0.0cm}
\centering
\includegraphics[width=0.24\textwidth]{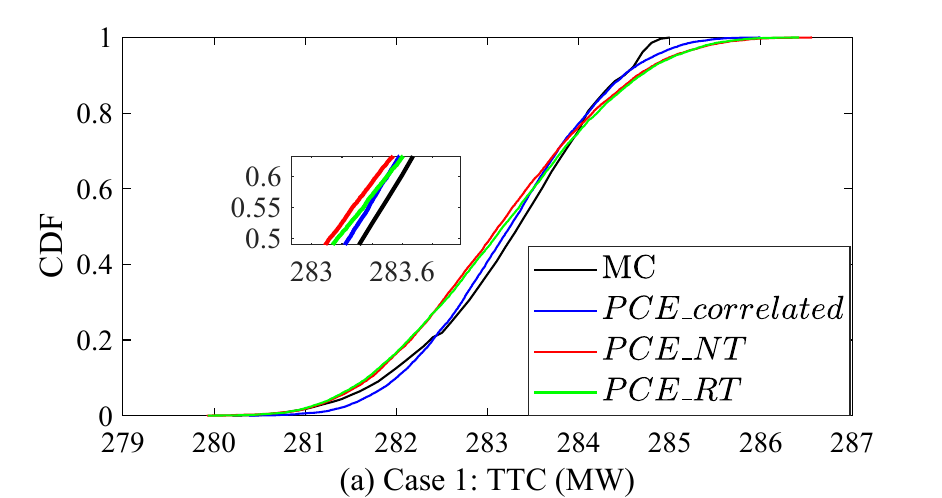}
\includegraphics[width=0.24\textwidth]{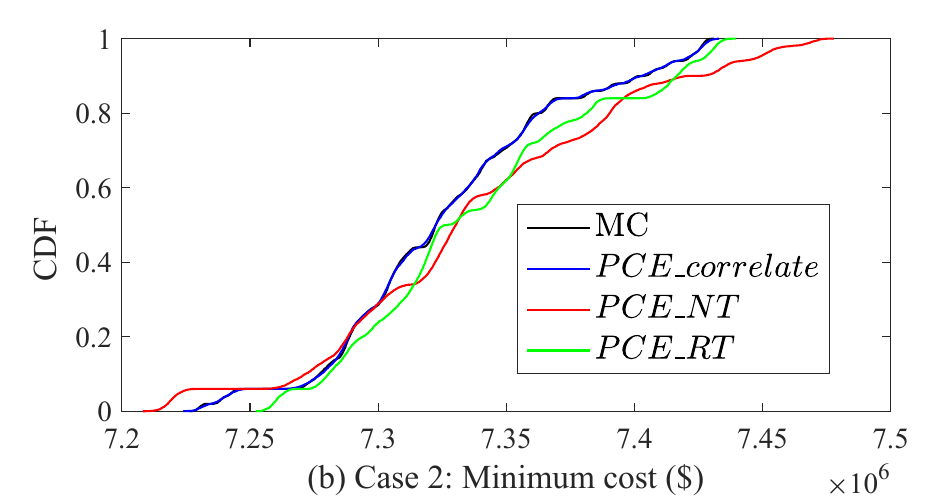}
\caption{Left (Case 1): the  CDFs of PTTC from the three PCE-based models and the MC simulations. Right (Case 2): the CDFs of the ED cost from the three PCE-based models and the MC simulations.} 
\label{fig:CDFPDF_Comp_case24} 
\end{figure}

\vspace{-0.07in}
We 
consider the IEEE 118-bus system integrated with a 20-node gas system 
\color{black} for the stochastic ED problem. Refer to \color{black}Section IV \cite{Wang2021ED} for the detailed system configuration\color{black}. There are $120$ random inputs (i.e., wind generator outputs), \color{black} and $Y$ is the ED cost. \color{black}  
Three PCE-based 
models are first constructed through $M_p = 1100$ sample pairs  (\textbf{Step 1-2}). Similarly, \color{black} the order $p$ determined by the LAR algorithm \cite{Wang2021} is 2 for all the PCE-based models. 
Next,  $M_L = 10,000$ correlated input samples are applied to each model (\textbf{Step 4})\color{black}. 
Fig. \ref{fig:CDFPDF_Comp_case24}-(b) 
shows that \textit{PCE\_correlate} provides the closest results to the benchmark MC simulation in estimating the CDF.

\begin{figure}[ht]
\vspace{-0.05in}
\setlength{\abovecaptionskip}{-0.21cm}
\setlength{\belowcaptionskip}{-0.3cm}
\centering
\includegraphics[width=0.37\textwidth]{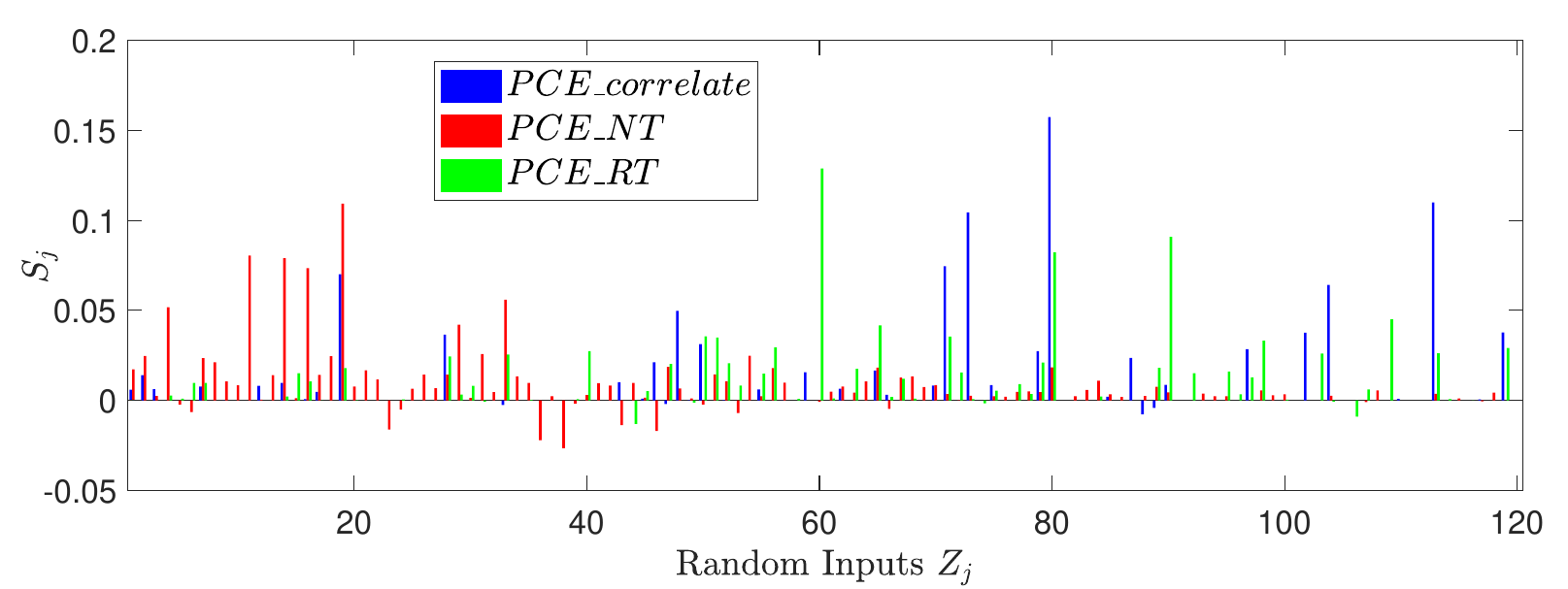}
\caption{
$S_j$ estimated by the three PCE-based models  
for the ED cost} 
\label{fig:ANCOVA_Case118}
\vspace{-0.06in}
\end{figure}

\begin{figure}[!ht]
\setlength{\abovecaptionskip}{-0.25cm}
\setlength{\belowcaptionskip}{-0.17cm}
\centering
\includegraphics[width=0.25\textwidth]{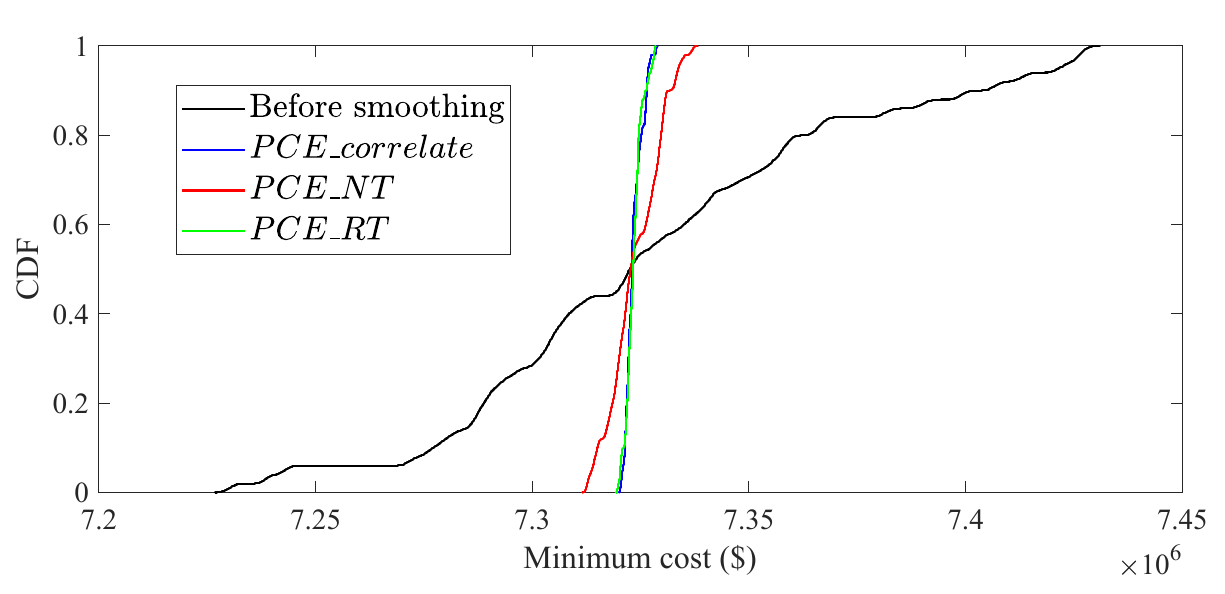}
\caption{The CDFs of the ED cost after smoothing the three sets of top 40 inputs by 
MC simulations.} 
\label{fig:CDFPDF_Comp_case118_smoothed}
\vspace{-0.04in}
\end{figure}

Next, the ANCOVA indices \small $S_j$ \normalsize are calculated (\small\textbf{Step 5}\normalsize) and ranked. \color{black} Fig. \ref{fig:ANCOVA_Case118} shows 
the top 40  dominant random inputs (a third 
of the total inputs) identified by the three PCE-based models. 
Surprisingly, the random inputs identified by the three PCE-based models are very different. For validation, the three  
sets of top 40 inputs are smoothed 
one by one by the three PCE-based model (\small\textbf{Step 6}\normalsize) and compared with MC simulations\color{black}. Table \ref{tab:statistic_comp_case118} shows that smoothing the top 40 dominant inputs  identified \color{black}
by \textit{PCE\_correlate} can reduce the variance of $Y$ most effectively ($\approx 95.9\%$ reduction \color{black} of the standard deviation of $Y$\color{black}) and accurately (closest to the MC simulations), while \textit{PCE\_RT} also gives reasonably accurate results.    
Fig. \ref{fig:CDFPDF_Comp_case118_smoothed}  shows the CDFs before and after smoothing the top 40 inputs identified by the three PCE-based models through the MC simulations. 
%
\subsection{Discussions of the Results} \label{sec:discussion}
\vspace{-0.06in}
In this section, we discuss the potential reasons for the \textit{first}  method provides better performance than the \textit{second}  method.\color{black} \\
\noindent\textbf{ The \textit{first method:}} Let \footnotesize $\rho_{j}(Z_j)$ 
 \normalsize be the marginal PDF  of \footnotesize $Z_j$ \normalsize and \footnotesize $\rho(\bm{Z})$ \normalsize be the joint PDF of \footnotesize$\bm{Z}$\normalsize. For dependent random inputs, 
to ensure the orthogonality of polynomial bases,  the multivariate orthogonal polynomial bases \footnotesize $\bm{P}_{\bm{k}}$  \normalsize should be built as (see \textit{Lemma 1} in \cite{Soize2004}): \footnotesize $\left(\prod_{j=1}^{D} \frac{\rho_{j}(Z_j)}{\rho(\bm{Z})} \right)^{\frac{1}{2}} \prod_{j=1}^{D}\phi^j_{k_j}(Z_{j})$ \normalsize  such that  
the convergence of the PCE model to the response \small $Y$ \normalsize in the sense of $L_2$ norm can be guaranteed. \color{black} 
However, \textit{PCE\_correlate} neglects the density term \footnotesize $\left[\prod_{j=1}^{D} \frac{\rho_{j}(Z_j)}{\rho(\bm{Z})} \right]^{\frac{1}{2}}$ \normalsize
and uses only \footnotesize $\prod_{j=1}^{D}\phi^j_{k_j}(Z_{j})$ \normalsize as the bases  \footnotesize $\bm{P}_{\bm{k}}$ \normalsize due to the fact that the joint PDF \footnotesize $\rho(\bm{Z})$ \normalsize \color{black} required by the method in \cite{Soize2004} \color{black} 
may not be obtained accurately and the  density term 
may be highly nonlinear in practice. Despite the inaccuracy introduced by neglecting the density term when establishing the polynomial basis, the calculation of \footnotesize$a_{\bm{k}}$ \normalsize  (\textbf{Step 2} c in \textbf{Method 1}) may compensate for the errors, still yielding relatively good accuracy, i.e., \footnotesize $Y\approx G^{\mathrm{pc}}(\bm{Z})$. \normalsize 

\noindent\textbf{The \textit{second} method:} \textit{PCE\_{NT}} or \textit{PCE\_{RT}} will first transform the random inputs \footnotesize $\bm{Z}_p$ \normalsize to \footnotesize $\bm{U}_p$\normalsize:  
\footnotesize $\bm{Z}_p =\mathcal{ T}^{-1}(\bm{U}_p)$\normalsize. 
Then, the PCE-based models are built through the sample pairs \footnotesize$[ \bm{U}_{{p}}, \bm{Y}_p]$\normalsize, i.e.,  
\footnotesize $Y\approx G^\mathrm{pc}(\bm{U}) =  \sum_{\bm{k}\in \mathbb{N}^{D} }a_{\bm{k}} P_{\bm{k}}(\bm{U}) $ \normalsize (\textbf{Step 2} in \textbf{Method 2}). 
Nevertheless, when evaluating ANCOVA indices in \textbf{Step 4} of \textbf{Method 2}, \footnotesize$\bm{U}$ \normalsize is replaced by  \footnotesize$\bm{Z}$ \normalsize in the established PCE-based models. I.e., we assume \footnotesize $Y\approx G^\mathrm{pc}(\bm{Z})$ \normalsize and use this model in calculating ANCOVA indices, even though the PCE-based models built in \textbf{Step 2} is to ensure \footnotesize $Y\approx G^\mathrm{pc}(\bm{U})=G^\mathrm{pc}(\mathcal{T}(\bm{Z}))$. \normalsize  
Errors are inevitably introduced due to the transformation \footnotesize $\mathcal{T}$\normalsize. Moreover, the errors cannot be compensated as the PCE-based models have already been built in \textbf{Step 2}. Hence, the \textit{second} method may underperform the \textit{first} method.    
\normalsize

%
\section{Conclusions}

We compared three PCE-based methods for global sensitivity analysis of correlated random inputs in two power system applications.  Simulation results have shown that \textit{PCE\_correlate}, which ignores input factor dependencies, provides the most accurate ANCOVA  
indices compared with  PCE-based models built after decorrelation using the Nataf transform or the Rosenblatt transform. Analysis of errors was also presented. Effective uncertainty control measures can be designed based on the ANCOVA
indices calculated by \textit{PCE\_correlate} to reduce the variance of system response and enhance the system performance. \color{black} Our future work involves considering only extremely small evaluations available in practice and combining the PCE methods with other metamodels (e.g., Kriging). Besides, 
\color{black} further analytical investigation will be carried out to comprehensively analyze and compare the performance of the three PCE-based models in global sensitivity analysis while considering different indices (e.g., ANCOVA, ANOVA).
\color{black} 
\vspace{-0.06in}
\color{black}


\end{document}